\documentclass[a4paper,11pt,hyper]{JHEP3}

\usepackage{amsmath,latexsym,amssymb}
\usepackage{graphicx}
\usepackage{psfrag}
\usepackage[latin1]{inputenc}
\usepackage{subfigure}
\usepackage{bbm}
\usepackage[stable]{footmisc}

\usepackage{comment}
\usepackage{color}

\newcommand\bbone{\ensuremath{\mathbbm{1}}}

\newcommand{\eq}[1]{\begin{equation}#1\end{equation}}
\newcommand{\spl}[1]{\begin{split}#1\end{split}}
\newcommand{\al}[1]{\begin{align}#1\end{align}}
\newcommand{\subeq}[1]{\begin{subequations}#1\end{subequations}}

\def\d{\text{d}}

\def\Re           {{\rm Re\hskip0.1em}}
\def\Im           {{\rm Im\hskip0.1em}}

\title{Coisotropic D-branes on ${\rm AdS}_4\times\mathbb{CP}^3$ and massive deformations}

\author{Paul Koerber\\
Max-Planck-Institut f\"ur Physik\\
F\"ohringer Ring 6, D-80805 M\"unchen, Germany\\

\bigskip
 E-mail:
\email{koerber at mppmu.mpg.de}}

\abstract{We scan for massive type IIA SU(3)-structure compactifications
of the type AdS$_4\times\mathbb{CP}^3$ with internal symmetry group SO(4).
This group acts on $\mathbb{CP}^3$ with cohomogeneity one, so that one would
expect new non-homogeneous solutions. We find however that all such
solutions enhance their symmetry group to Sp(2) and form, in fact, the homogeneous
family first described in \cite{tomasiellocosets}. This is in accordance with
\cite{tomasiellomassive}, which argues from the CFT-side that although new
vacua with SO(4) symmetry group and $N=2$ supersymmetry should exist, they
fall outside our ansatz of strict SU(3)-structure, and instead have
genuine SU(3)$\times$SU(3)-structure. We do find that the SO(4)-invariant description,
which singles out one preferential direction in the internal space, is
well-adapted for describing the embedding of AdS$_4$-filling supersymmetric D8-branes
on both the original ABJM configuration as its massive Sp(2)-symmetric deformations.
Supersymmetry requires these D-branes to be of the coisotropic type, which means in
particular that their world-volume gauge field must be non-trivial.
}

\preprint{MPP-2009-35}

\begin{document}
\setcounter{footnote}{0}
\renewcommand{\thefootnote}{\arabic{footnote}}
\setcounter{section}{0}
\section{Introduction}
\label{introduction}

Supersymmetric compactifications with fluxes have become an important ingredient
in the search for realistic vacua of string theory, since some or even all of the moduli can be stabilized (for reviews
see e.g.~\cite{fluxrev1,fluxrev2,fluxrev3,fluxrev4}). Apart from phenomenological reasons, the philosophy
behind imposing supersymmetry is that solving for the supersymmetry conditions is much easier than solving
all the equations of motion, while it has been shown that for (eleven-dimensional and type II) supergravity these supersymmetry conditions completed with the Bianchi identities indeed imply the equations of motion \cite{gauntlettM,lt,gauntlettIIB,integr}. To obtain solutions with
a positive cosmological constant, however, one has to break supersymmetry. A further
complication for constructing compactifications with a non-negative four-dimensional cosmological constant is that in the presence of fluxes the no-go theorem of \cite{malnun} requires the introduction of negative-tension sources. Although in string theory they can be provided for by
orientifolds, these localized sources complicate the explicit form of the solution. The strategy is then to construct
solutions with a negative cosmological constant, with AdS$_4$ as the external space, and
take care of both uplifting to dS$_4$ and supersymmetry breaking in a later stage.

Because of the gauge/gravity correspondence type IIA solutions with an AdS$_4$
factor are also interesting as geometries potentially dual to a three-dimensional conformal theory. Following
\cite{abjm}, where the three-dimensional CFT is a Chern-Simons-matter theory, there has been lot of progress in this
direction recently. In this way these solutions can inspire the construction of new three-dimensional CFTs or, conversely, the CFTs
can lead the way to the construction of new geometries.

The supersymmetry and Bianchi conditions for type IIA AdS$_4$ flux solutions with strict SU(3)-structure were first
worked out in \cite{lt}.\footnote{I use here the terminology {\em strict} SU(3)-structure for a supersymmetry
ansatz where the same globally-defined internal spinor enters in both the left- and right-moving supersymmetry generator of
type II supergravity, in order to distinguish it from the more general SU(3)$\times$SU(3)-structure ansatz, where
two different internal spinors enter.} They lead to very specific requirements
on the internal geometry. In particular, only the SU(3)-structure torsion classes $\mathcal{W}_1$ and $\mathcal{W}_2$
can be non-zero and they can be chosen to be purely imaginary.

Prominent examples of solutions to these equations are homogeneous constructions on coset manifolds
\cite{cveticnk1,cveticnk2,housepalti,font,tomasiellocosets,cosets}. In particular, in \cite{tomasiellocosets}
a family with one shape parameter was constructed on $\mathbb{CP}^3=\frac{\text{Sp(2)}}{\text{S}(\text{U(2)}\times \text{U(1)})}$.
Changing the shape parameter, these Sp(2)-invariant solutions interpolate between the solution of \cite{nilssonpope} -- with vanishing
Romans mass and standard Fubini-Study metric -- and a second, squashed, massless solution. While the two endpoints of the family can be lifted
to M-theory and correspond to the standard $S^7$ and the squashed $S^7$-solution respectively, the rest of the family has non-zero Romans mass and includes
in particular a nearly-K\"ahler solution (i.e.\ $\mathcal{W}_2=0$). It was argued that a similar story applies to $\frac{\text{SU(3)}}{\text{U(1)}\times \text{U(1)}}$, although it was later found \cite{cosets}  that these SU(3)-invariant solutions actually have two shape parameters.
The effective four-dimensional theory corresponding to an expansion in terms of left-invariant forms on these cosets was studied in
\cite{effectivecosets,effectivekoers}. It was found in \cite{cassred} that this expansion actually corresponds to a consistent reduction.

The generic solution in the family has $N=1$ supersymmetry. However, for the massless solution on $\mathbb{CP}^3$ with Fubini-Study
metric, the supersymmetry enhances to $N=6$. At the same time the bosonic symmetry group enhances from Sp(2) to SU(4) and corresponds to
the description of $\mathbb{CP}^3$ as the coset manifold $\frac{\text{SU(4)}}{\text{S(U(3)}\times\text{U(1))}}$. It got recently a lot of attention as
the dual geometry for the original ABJM gauge/gravity correspondence \cite{abjm} in
the limit where the type IIA description is valid. The geometry then also has a K\"ahler form and becomes Einstein, so we will dub it
the massless K\"ahler-Einstein solution in the following. Also for the squashed massless case, a CFT dual has in the meantime been proposed \cite{oogurisquashed}.

A natural question that arises then, namely whether also CFT duals can be found for the interpolating solutions with non-zero Romans mass, was addressed in \cite{tomasiellomassive} (see also \cite{massiveABJM}). Surprisingly, not only these could be found, but
in fact several massive deformations of the ABJM Chern-Simons-matter theory were proposed with different numbers
of supersymmetries and different global symmetry groups: $N=0$ with SO(6)-symmetry, $N=1$ with Sp(2) symmetry, $N=2$ with SO(4)$\times$SO(2)$_R$ and
finally $N=3$ with SO(3)$\times$SO(3)$_R$. It was argued that the $N=0$ theories correspond to AdS$_4$ solutions with
the standard Fubini-Study metric on $\mathbb{CP}^3$, but with different fluxes (and in particular of course non-zero Romans mass) from the ones in the $N=6$ configuration. Similar non-supersymmetric AdS$_4$ vacua were constructed in \cite{gennonsusy,dimitriosextrasol}, and it would be interesting to investigate whether they also allow for CFT duals. The second family with $N=1$ supersymmetry and Sp(2) bosonic symmetry group then corresponds to the dual of the above interpolating solutions, while we come back to the last two families in a minute.

In this paper I will scan for supersymmetric solutions on $\mathbb{CP}^3$ with strict SU(3)-structure and a SO(4) symmetry group,
embedded in SU(4) such that it acts with cohomogeneity one on $\mathbb{CP}^3$.\footnote{\label{embfoot} There is another inequivalent
embedding of SO(4) in SU(4), which acts with cohomogeneity three. This is the symmetry group of the Lagrangian
D-branes of \cite{slagCP3,slagCP32,slagCP33}. The difference can also be seen on how it acts on the $\bf{6}$ (the antisymmetric two-tensor) of SU(4),
the representation under which the supersymmetry generators transform. Under the first SO(4)=SU(2)$\times$SU(2) -- the one considered in this paper -- it decomposes as $\bf{6}\rightarrow (\bf{1},\bf{1}) \oplus (\bf{1},\bf{1}) \oplus (\bf{2},\bf{1}) \oplus (\bf{1},\bf{2})$, and there are two supersymmetry generators singled out corresponding to the two first terms in the decomposition. For the second SO(4) we find $\bf{6} \rightarrow (\bf{3},\bf{1}) \oplus (\bf{1},\bf{3})$. The $N=3$ supersymmetry preserved by a Lagrangian D6-brane correspond then to the first term.} With only one transversal coordinate
this would provide the easiest setup for constructing non-homogeneous solutions. Unfortunately, we find that all such solutions
enhance their symmetry group to Sp(2) and are in fact the known family of $N=1$ solutions mentioned before, in disguise.

This is in agreement with the results of \cite{tomasiellomassive}, where 
although the third family of three-dimensional Chern-Simons-matter theories presented there has SO(4) global symmetry, it also has
$N=2$ supersymmetry and SO(2) R-symmetry. It was then argued that the geometric dual to this CFT could not be described
with a strict SU(3)-structure ansatz, but would instead have two genuine SU(3)$\times$SU(3)-structures.
The reason is that the supersymmetry conditions of \cite{lt} directly relate the (3,0)-form $\Omega$ with the NSNS-flux $H$. So if both supersymmetries would be of the strict SU(3)-structure type, they would correspond to the same $\Omega$ and thus the same almost complex structure. Since from the metric and the almost complex structure one can then also find the two-form $J$, the SU(3)-structures and thus supersymmetries would in the end be entirely the same. Furthermore, since the R-symmetry requires
both supersymmetries to be of the same type, they must be both SU(3)$\times$SU(3).
 The only way around the argument is for $m=0$ where also $H=0$ and there is no relation anymore between $\Omega$ and $H$. This loophole makes it possible for the standard ABJM $\mathbb{CP}^3$ to be $N=6$ and still have strict SU(3)-structure. In the meantime the geometries with SU(3)$\times$SU(3)-structure corresponding to these $N=2$ (and also $N=3$) CFTs were constructed in \cite{tomasiellomassive2} to first order in the mass parameter. A similar geometry on $M^{1,1,1}$ was obtained
in \cite{petrinimassive}.

The SO(4)-invariant setup is however useful to describe supersymmetric AdS$_4$-filling D8-branes. Indeed, these D-branes
will have one transversal direction and wrap the five-dimensional internal cycle on which the SO(4) acts homogeneously.
According to \cite{gencal,lucacal} the supersymmetry conditions for D-branes force them to be generalized calibrated. Just as for the supergravity background,
they are also sufficient to ensure that the D-brane solves its equations of motion. In the case of AdS$_4$ compactifications there are
some additional subtleties that were studied in detail in \cite{DbraneAdS}. One interesting aspect is, for instance, that the so-called D-flatness condition will follow automatically from the F-flatness conditions. For space-time-filling D6-branes, for example, that means that a Lagrangian cycle will automatically be special (with respect to the SU(3)-structure (J,$\Omega$) associated to the bulk supersymmetry). For the particular case of space-time-filling D8-branes, the supersymmetry conditions imply that they must be coisotropic \cite{coisotropic}, which means in particular they must have a non-trivial world-volume gauge field. Examples of coisotropic D-branes are still quite rare: there are constructions on the torus (see e.g.~\cite{coisotropicmarchesano} for an application) and in the context of the world-sheet approach (see e.g.\ \cite{wijnscoisotropic}). In the latter case it is however not yet clear whether the D-flatness condition can be satisfied. Coisotropic D-branes (on the Hitchin moduli space) also played role in the construction of \cite{wittenlanglands}. In this paper, I construct coisotropic embeddings on the whole
family of $N=1$ Sp(2)-invariant solutions. In the K\"ahler-Einstein limit these coisotropic D-branes will preserve $N=2$, while away from the limit they preserve
the $N=1$ of the bulk.

In section \ref{structure} I study the constraints on the solutions from having the SO(4) symmetry group with the embedding in SU(4) discussed above. I find that the solution should have cohomogeneity one, and that the five-dimensional orbits of SO(4) form the coset $T^{1,-1}=\frac{\text{SO(4)}}{\text{U(1)}}$, better known as the base of the conifold \cite{conifolds}.
I rewrite the starting point, namely the $N=6$ massless solution with Fubini-Study metric, in terms of coordinates reflecting the
SO(4)-symmetry. In section \ref{condreview} I review the conditions for strict SU(3)-structure of \cite{lt}, which I solve in section \ref{condsol}, starting from the ansatz suggested by the symmetry analysis of section \ref{structure}. In section \ref{proofequiv}, I show that the single resulting family of solutions is actually equivalent to the known family of Sp(2)-invariant solutions. In section \ref{coisemb} I construct coisotropic D-brane embeddings on this family.

\section{Setup: SO(4)-invariant ansatz}
\label{structure}

In this section I analyse the constraints that the SO(4) symmetry group
imposes on the metric and the form fields of the solution.
It turns out that the SO(4) symmetry is not large enough to make the solutions homogeneous.
Instead, they have cohomogeneity one, which is the next easiest case. The ansatz for
the metric and the form fields will then depend on functions in only one variable.

\subsection{Symmetry group and codimension-one foliation}
\label{symmetries}

The ABJM theory \cite{abjm} consists of a U$(N)_k\times$U$(N)_{-k}$ Chern-Simons theory
with chiral multiplets
\eq{
\label{ABtrans}
(A_1,A_2,B_1^*,B_2^*)
}
transforming as a $\bf{4}$ under the global symmetry group SU(4). According to \cite{tomasiellomassive}
turning on the Romans mass, in the $N=2$ branch this global symmetry is broken to SO(4)$\times$SO(2)$_R$. The
SO(4)=SU(2)$_A\times$SU(2)$_B$ part acts such that
\eq{
(A_1,A_2) \, , \qquad (B_1,B_2)
}
transform as $\bf{2}_A$ and $\bf{2}_B$ respectively.

The dual of the massless ABJM theory in the regime where type IIA is valid,
is the AdS$_4\times\mathbb{CP}^3$ solution of \cite{nilssonpope}. The internal space
$\mathbb{CP}^3$ is then equipped with the Fubini-Study metric and is described by the
coset
\eq{
\label{cosetSU4}
\frac{\text{SU(4)}}{\text{S(U(3)}\times\text{U(1))}} \, .
}
The homogenous coordinates of $\mathbb{CP}^3$,
\eq{
(Z^1,Z^2,Z^3,Z^4) \cong \lambda (Z^1,Z^2,Z^3,Z^4) \, , \quad \text{with} \,\, \lambda\in\mathbb{C}_* \, ,
}
then transform as a $\bf{4}$ just like the $(A_i,B_i^*)$ in eq.~\eqref{ABtrans}. We therefore deduce that for the
deformed geometry the isometry SO(4) should act such that
\eq{
\label{so4action}
(Z^1,Z^2) \, , \qquad (Z^3,Z^4)
}
transform as $\bf{2}_A$ and $\bf{2}_B$ respectively. This particular SO(4)-embedding thus breaks
$\bf{4} \rightarrow (\bf{2},\bf{1}) \oplus (\bf{1},\bf{2})$ and as we will discuss in section
\ref{proofequiv} also $\bf{6} \rightarrow (\bf{1},\bf{1}) \oplus (\bf{1},\bf{1}) \oplus (\bf{2},\bf{1}) \oplus (\bf{1},\bf{2})$.
As discussed in footnote \ref{embfoot} and as we will show presently, it is this embedding of SO(4) that acts with
cohomogeneity one on the internal space.

So let us study the orbits of SO(4). First we turn to special points with coordinates of the form
\eq{
\label{specialpoints1}
(Z^1,Z^2,0,0) \, .
}
One can easily check that they are rotated into each other by $SU(2)_A$ and that the part
of the original isotropy group $\text{S(U(3)}\times\text{U(1))}$ that belongs
to SO(4) is $U(1)\times SU(2)_B$. The SU(2)$_B$ factor cancels out in numerator and denominator so that the orbit is
\eq{
\mathbb{CP}^1_A = \{(Z^1,Z^2) : (Z^1,Z^2) \cong \lambda (Z^1,Z^2) \} = \frac{\text{SU(2)}_A}{\text{U(1)}} \, .
}
Likewise, the orbit of the points of the special form
\eq{
\label{specialpoints2}
(0,0,Z^3,Z^4)
}
is also a $\mathbb{CP}^1$, which we call $\mathbb{CP}^1_B$.

Let us now turn to the
orbit through a point with coordinates of the form
\eq{
(0,Z^2,Z^3,0) \, .
}
Since acting with SO(4) one can bring a generic point in this form, this is the generic case.
We find that this time the part of the isotropy group $\text{S(U(3)}\times\text{U(1))}$ that belongs to SO(4)
is a U(1) acting like
\eq{
\label{actionU1}
(Z^1,Z^2,Z^3,Z^4) \rightarrow (e^{i\phi}Z^1,e^{-i\phi}Z^2,e^{-i\phi}Z^3,e^{i\phi}Z^4) \, .
}
We conclude that the orbit of SO(4) through this point is five-dimensional and
given by the coset manifold
\eq{
\frac{\text{SO(4)}}{\text{U(1)}} = T^{1,-1} \, ,
}
which is better known as the base of the conifold (see e.g.~\cite{conifolds}). It has the topology of $S^2 \times S^3$.
$\mathbb{CP}^3$ can thus locally be considered as a foliation with the generic leaves taking the form of the coset $T^{1,-1}$.
This is not globally a foliation though, since this description breaks down at special points of the
forms \eqref{specialpoints1} and \eqref{specialpoints2}, where the leaves collapse to two-dimensional $\mathbb{CP}^1$s.
When constructing metrics on $\mathbb{CP}^3$ in this description it is therefore important to carefully check the regularity on these degenerate orbits.
That $\mathbb{CP}^3$ could be locally seen as a  foliation in this way was noted before in \cite{CP3foliation}.

\begin{figure}[tp]
\centering
\definecolor{darkblue}{rgb}{0,0,0.5}
\definecolor{darkgreen}{rgb}{0,0.5,0}
\definecolor{darkred}{rgb}{0.82,0,0}

\psfrag{U1}{\color{darkgreen}U(1)}
\psfrag{CPA}{\color{darkblue}$\mathbb{CP}^1_A$}
\psfrag{CPB}{\color{darkred}$\mathbb{CP}^1_B$}
\psfrag{xi0}{$t=0$}
\psfrag{xipi}{$t\rightarrow+\infty$}
\includegraphics[width=10cm]{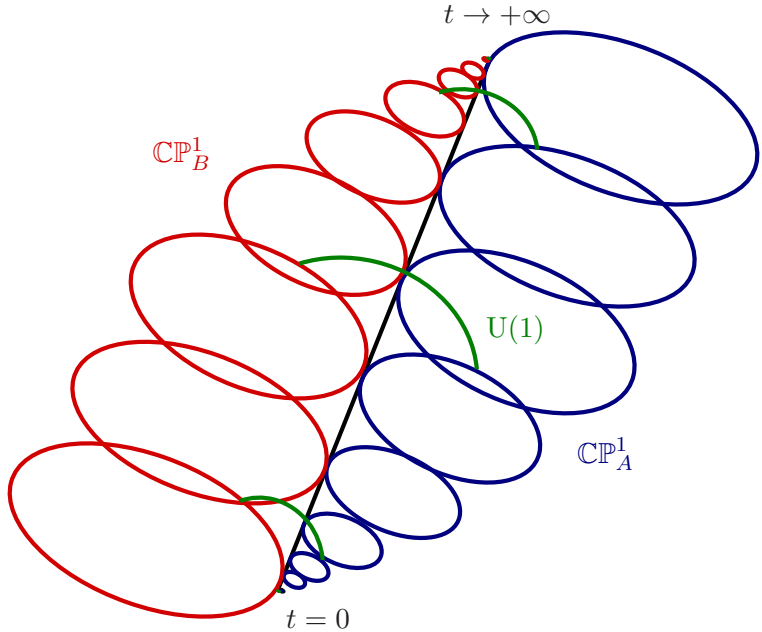}
\caption{Representation of the geometry of the solutions on $\mathbb{CP}^3$. As explained in the text, the
coset $T^{1,-1}$ can be seen as a U(1)-bundle over $\mathbb{CP}^1_A \times \mathbb{CP}^1_B$. How the sizes
of these $\mathbb{CP}^1$s vary with $t$ depends on the solution and is described
by the metric \protect\eqref{fbmetric}
in the K\"ahler-Einstein case or by the metric \protect\eqref{genmetric} for the general solution. In all cases the U(1) and
$\mathbb{CP}^1_{A},\mathbb{CP}^1_{B}$ shrink to zero for $t \rightarrow 0,+\infty$ respectively.
}
\label{geometrystruc}
\end{figure}

Another way to see this foliation, which allows for a better handle on the location of the two special loci, is
to start from a generic point $(Z^1,Z^2,Z^3,Z^4)$ and note that the action
of SO(4)=SU(2)$_A \times$SU(2)$_B$ allows to go to any other point with
the same absolute values $|Z^1|^2+|Z^2|^2$ and $|Z^3|^2+|Z^4|^2$.
Since these are homogenous coordinates, there is only meaning to the relative factor $t=(|Z^1|^2+|Z^2|^2)/(|Z^3|^2+|Z^4|^2)$.
This implies that the orbit of the point $(Z^1,Z^2,Z^3,Z^4)$ contains all the points with the same relative factor $t$.
The two special orbits then correspond to sending the factor $t$ to zero and infinity  respectively.
The first two coordinates can be seen to describe a $\mathbb{CP}^1=S^2$, for which we can choose
inhomogeneous coordinates fixing the scaling freedom. For the last two coordinates the absolute value $|Z^3|^2+|Z^4|^2$ is then fixed
so that we find an $S^3$. In the special loci, the radius of this three-sphere or the one with the roles of $(Z^1,Z^2)$ and $(Z^3,Z^4)$
interchanged goes to zero. See figure \ref{geometrystruc} for a an illustration of the geometry.

The space $T^{1,-1}$ can also be considered as a U(1)-bundle over $\mathbb{CP}^1_A \times \mathbb{CP}^1_B$.
Indeed the two $\mathbb{CP}^1$s are described by the homogeneous coordinates $(Z^1,Z^2)$ and $(Z^3,Z^4)$
respectively, and the U(1) describes the relative phase between the two factors.

It will turn out that in the massless K\"ahler-Einstein limit the SO(2)$_R$ symmetry that rotates both SO(4)-invariant
supersymmetries into each other, acts as follows
\eq{
\label{Rsymmetry}
(Z^1,Z^2) \rightarrow e^{i \phi/2} (Z^1,Z^2) \, , \qquad (Z^3,Z^4) \rightarrow e^{-i\phi/2} (Z^3,Z^4) \, .
}

Let us now study the coset $T^{1,-1}$, which makes up the geometry of the five-dimensional orbits, in
a bit more detail.

\subsection{The coset $T^{1,-1}$}

Imposing that our solutions respect the SO(4) symmetry group, implies that the metric, the SU(3)-structure
and the form fields should all be invariant under the left action of this group. In this section we construct the most general
form of the SO(4) left-invariant metric and forms on the $T^{p,q}$ cosets. For a review on
coset manifolds and their description in terms of left-invariant forms see \cite{cosetreview1,cosetreview2}.
For an overview of the $T^{p,q}$ coset in particular -- which also explains its description as
a U(1) bundle over $\mathbb{CP}^1 \times \mathbb{CP}^1$ -- see \cite{conifolds}

As explained in these reviews (and also in \cite{cosets,effectivecosets}) on a coset manifold $G/H$ we must split off the generators $H_a$
of the algebra corresponding to the group $H$ that is projected out from the rest of the generators $E_i$ of $G$.
This induces a split of the structure constants:
\eq{\spl{\label{commut}
[H_a,H_b]&=f^c{}_{ab}H_c \, , \\
[H_a,K_i]&=f^j{}_{ai}K_j+f^b{}_{ai}H_b \, , \\
[K_i,K_j]&=f^k{}_{ij}K_k+f^a{}_{ij}H_a \, ,
}}
where for a reductive coset, which is the case for us, the split basis can be chosen such that $f^b{}_{ai}=0$.
The decomposition of the Lie-algebra valued one-form
\eq{
L^{-1}\d L= e^i K_i+\omega^a H_a \, ,
}
for $L \in G$, defines then a coframe $e^i(y)$ on $G/H$. The left-invariant $l$-forms $\phi$ are then the
forms which can be expanded in terms of the coframe $e^i$ with coefficients constant over the coset,
and moreover satisfy
\eq{
\label{leftinvform}
f^j{}_{a[i_1}\phi_{i_2\dots i_l]j}=0 \, .
}
In our case however, we must allow the coefficients to depend on the one coordinate transversal to the coset.
Likewise a left-invariant metric must have coefficients constant over the coset in the coframe basis $e^i$
and satisfy the condition
\eq{
f^k{}_{a(i} g_{j)k}=0 \, .
\label{leftinvmetric}
}
The exterior derivatives of the one-forms of the coframe can be expressed in terms
of the structure constants as follows:
\eq{
\d e^i=-\frac{1}{2}f^i{}_{jk} e^j\wedge e^k - f^i{}_{aj} \omega^a\wedge e^j \, .
}
Acting with the exterior derivative on left-invariant forms, the condition \eqref{leftinvform} will
ensure that all terms containing $\omega^a$ drop out.

Suppose the SU(2)$_{A/B}$ factors in SO(4) are generated by $\lambda_{A,B}= -\frac{i}{2} \sigma_{A,B}$, where
$\sigma_{A,B}$ are the Pauli matrices, then we propose the following split for the generators of the algebra of SO(4)
\eq{\spl{
(E_1,\ldots,E_5) & = (\lambda_{A1},\lambda_{A2},\lambda_{B1},\lambda_{B2},\frac{1}{2p} \lambda_{3A} +\frac{1}{2q} \lambda_{3B})\, ,\\
H_1 & = \frac{1}{2p} \lambda_{3A} -\frac{1}{2q} \lambda_{3B} \, .
}}
Here $H$ is the generator of the U(1) that is projected out.

This leads to the structure constants:
\eq{\label{struccons}\spl{
& f^5{}_{12}=f^6{}_{12} = p \, , \quad f^5{}_{34} = -f^6{}_{34} =q \, ,
\quad f^1{}_{25} = f^2{}_{51} = \frac{1}{2p} \, , \quad f^3{}_{45} = f^4{}_{53} = \frac{1}{2q} \, , \\
& f^1{}_{26} = f^2{}_{61} = \frac{1}{2p} \, , \quad f^3{}_{46} = f^4{}_{63} = - \frac{1}{2q} \, .
}}

If $p \neq q$ one can choose the following coordinate representation\footnote{For $p=q$ this particular representation is not so great,
since the terms with $\d\psi$ drop out of $e^5$. For this case another more suitable coordinate representation can be found by changing some signs. Here we are
interested in the case $p=-q=1$ anyway.} for the one-forms $(e^i,\omega^a)$:
\eq{\label{Npqcoordrep}\spl{
e^1 & = - \sin \psi \, \d \theta_1 + \sin \theta_1 \cos \psi \, \d \phi_1  \, , \\
e^2 & = \cos \psi \, \d \theta_1 + \sin \theta_1 \sin \psi \, \d \phi_1  \, , \\
e^3 & = -\sin \psi \, \d \theta_2 - \sin \theta_2 \cos \psi \, \d \phi_2  \, , \\
e^4 & = -\cos \psi  \, \d \theta_2 + \sin \theta_2 \sin \psi \, \d \phi_2  \, , \\
e^5 & = - \left[(p-q) \, \d \psi + p \cos \theta_1 \, \d \phi_1 + q \cos \theta_2 \, \d \phi_2\right] \, , \\
\omega^1 & = - \left[(p+q) \, \d \psi + p \cos \theta_1 \, \d \phi_1 - q \cos \theta_2 \, \d \phi_2 \right] \, ,
}}
where $(\theta_{1,2},\phi_{1,2})$ are spherical coordinates on $\mathbb{CP}^1_{A,B}$ respectively,
satisfying $0 \le \theta_{1,2} < \pi$, $0 \le \phi_{1,2} < 2 \pi$, and $\psi$ describes the U(1)-bundle, with $0 \le \psi < 2 \pi$.

From eq.~\eqref{actionU1} describing the action of the U(1) we find that the case of interest is $p=-q=1$.
One finds then from \eqref{leftinvmetric} that in the coframe basis $e^i$, a left-invariant metric should satisfy
\eq{
g_{11} = g_{22} = m_1 \, ,\quad g_{33} = g_{44} = m_2 \, , \quad g_{55}=m_3 \, , \quad g_{13} = g_{24} = m_4 \, , \quad g_{14}= - g_{23} = m_5 \, ,
}
with $(m_1,m_2,m_3,m_4,m_5)$ five functions of the transversal coordinate. In the explicit representation \eqref{Npqcoordrep} the metric becomes
\begin{multline}
\label{N11metric}
ds^2 =  m_1 \left[ (\d \theta_1)^2 + \sin^2 \theta_1 (\d \phi_1)^2 \right] +
m_2 \left[ (\d \theta_2)^2 + \sin^2 \theta_2 (\d \phi_2)^2 \right] + \\
 m_3 \left[2 \d \psi + \, \cos \theta_1 \d \phi_1 - \, \cos \theta_2 \d \phi_2 \right]^2 \hspace{5cm}\\
 + (- m_4 \cos 2 \psi+ m_5 \sin 2 \psi) (\d \theta_1 \d \theta_2 + \sin \theta_1 \sin \theta_2 \d \phi_1 \d \phi_2)\\
 +(m_4 \sin 2 \psi + m_5 \cos 2 \psi)(\sin \theta_2 \d \theta_1 \d \phi_2 - \sin \theta_1 \d \theta_2 \d \phi_1) \, .
\end{multline}

Furthermore, from \eqref{leftinvform} one finds that the left-invariant forms are spanned by
\eq{\label{leftinvforms}\spl{
1-\text{forms} & : e^5 \, , \\
2-\text{forms} & : e^{12}, e^{34}, e^{14}-e^{23}, e^{13} + e^{24} \, , \\
3-\text{forms} & : e^{125}, e^{345}, e^{145}-e^{235}, e^{135}+e^{245} \, , \\
4-\text{forms} & : e^{1234} \, , \\
5-\text{forms} & : e^{12345} \, .
}}

Finally, we remark that the structure constants \eqref{struccons} are invariant under
a shift of $\psi$:
\eq{
\label{psishift}
\psi \rightarrow \psi + \phi(t) \, ,
}
where we can allow the shift to depend explicitly on the transversal coordinate $t$.
This implies that the conditions we will impose on our solutions in section \ref{conditions} are invariant under this shift,
which will thus send SO(4)-invariant solutions to SO(4)-invariant solutions. If $\phi$ is constant,
the shift reduces to the SO(2)$_R$ symmetry \eqref{Rsymmetry}. The non-constant part we will appropriately
gauge away.

\subsection{Starting point: the massless K\"ahler-Einstein geometry}
\label{m0case}

The starting point of the analysis is the $N=6$ SU(4)-invariant vacuum on $\mathbb{CP}^3$ with
the standard Fubini-Study metric in terms of the homogeneous coordinates $(Z^1,Z^2,Z^3,Z^4)$.
I will rewrite this vacuum
in terms of the coordinates induced by the local foliation with $T^{1,-1}$ leaves.
Only the SO(4) isometry group and $N=2$ supersymmetry are then manifest.

Corresponding to the description of $T^{1,-1}$ as a U(1)-bundle over $\mathbb{CP}^1 \times \mathbb{CP}^1$,
appropriate coordinates are
\eq{
z = \frac{Z^1}{Z^2}  \, , \qquad w = \frac{Z^3}{Z^4}  \, , \qquad \psi = \frac{1}{2} \, \text{arg} \left(\frac{Z^1Z^2}{Z^3Z^4}\right) \, ,
}
where $z$ and $w$ are inhomogeneous coordinates for $\mathbb{CP}^1_{A,B}$ respectively.
Furthermore it will be convenient to choose a transversal coordinate $\xi$ as follows
\eq{
\tan^2 \xi = t = \frac{|Z^1|^2 + |Z^2|^2}{|Z^3|^2+|Z^4|^2} \, ,
}
and thus $0 \le \xi \le \frac{\pi}{2}$. If we further replace the inhomogeneous coordinates $(z,w)$
for both $\mathbb{CP}^1$s by the respective spherical coordinates $(\theta_{1,2},\phi_{1,2})$
we find in the end:
\eq{\label{homcoord}\spl{
Z^1 & = \lambda \sin \xi \cos \frac{\theta_1}{2} \exp \frac{i}{2} \left( \psi + \phi_1 \right) \, , \\
Z^2 & = \lambda \sin \xi \sin \frac{\theta_1}{2} \exp \frac{i}{2} \left( \psi - \phi_1 \right) \, , \\
Z^3 & = \lambda \cos \xi \cos \frac{\theta_2}{2} \exp \frac{i}{2} \left(-\psi + \phi_2 \right) \, , \\
Z^4 & = \lambda \cos \xi \sin \frac{\theta_2}{2} \exp \frac{i}{2} \left(-\psi - \phi_2 \right) \, ,
}}
where $\lambda$ is the inessential overall complex factor of the homogeneous coordinates.

The Fubini-Study metric becomes in these coordinates:
\begin{multline}
\label{fbmetric}
a^{-1} \d s^2 =  \d \xi^2 + \frac{\sin^2 \xi}{4} \left[ (\d \theta_1)^2 + \sin^2 \theta_1 (\d \phi_1)^2 \right]
+ \frac{\cos^2 \xi}{4} \left[ (\d \theta_2)^2 + \sin^2 \theta_2 (\d \phi_2)^2 \right] \\
+ \sin^2 \xi \cos^2 \xi \, \left(\d \psi + \frac{1}{2} \cos \theta_1 \d \phi_1 - \frac{1}{2} \cos \theta_2 \d \phi_2\right)^2
\, ,
\end{multline}
where we introduced a constant overall scale $a>0$.

Note that the orbits indeed correspond to $T^{p,q}$ with $p=-q=1$, and furthermore to putting
\eq{
m_1 = \frac{\sin^2 \xi}{4}\, , \qquad m_2 = \frac{\cos^2 \xi}{4} \, , \qquad m_3=\frac{\sin^2 \xi \cos^2 \xi}{4} \, , \qquad m_4=m_5=0 \, ,
}
in the coset metric of \eqref{N11metric}.
The transversal coordinate $\xi$ is chosen so that $g_{\xi\xi}=a$ is constant. These coordinates for $\mathbb{CP}^3$ were already
proposed in \cite{CP3foliation}.

Since it is just the standard Fubini-Study metric in new coordinates, we know that this metric is regular on the degenerate orbits $\xi \rightarrow 0$ and $\xi \rightarrow \pi/2$, where respectively the first and the second
$\mathbb{CP}^1$ as well as the fiber collapse. It is a useful
exercise to check this explicitly, since it shows us what to pay attention to when imposing the boundary conditions on the general solutions. In the limit $\xi \rightarrow 0$
the metric becomes
\begin{multline}
\label{metricreg}
a^{-1} \d s^2 = \d \xi^2 + \frac{\xi^2}{4} \left[ (\d \theta_1)^2 + \sin^2 \theta_1 (\d \phi_1)^2 + \left(2 \d \psi + \cos \theta_1 \d \phi_1 - \cos \theta_2 \d \phi_2\right)^2 \right] \\
+ \frac{1}{4} \left[ (\d \theta_2)^2 + \sin^2 \theta_2 (\d \phi_2)^2 \right]
\, .
\end{multline}
For constant $(\xi,\theta_2,\phi_2)$ the second term in the first line becomes the standard metric of an $S^3$.
Checking the volume of this $S^3(\xi)$ we find:
\eq{
\text{vol}(S^3(\xi)) = \left(\frac{\xi}{2}\right)^3 \int \sin \theta_1 \d \theta_1 \d \phi_1 2 \d \psi = (2 \pi^2) \xi^2 \, ,
}
the standard volume of an $S^3$ with radius $\xi$. For this calculation the prefactor $1/4$ in front of the second term in the first line, together with the
period of $\psi$ is crucial. It follows that the part of the metric in the first line is just the flat $\mathbb{R}^4$ metric, and thus regular.
In a completely analogous way, the metric is also regular for $\xi \rightarrow \pi/2$.

The standard closed K\"ahler form $J'$ is in these coordinates given by
\eq{
\label{FSKahl}
a^{-1} J' = \frac{\sin^2 \xi}{4} e^{12} + \frac{\cos^2 \xi}{4} e^{34} + \frac{\sin 2\,\xi}{4} e^5 \wedge \d \xi \, .
}
Note that this is {\em not} the two-form $J$ of the SU(3)-structure associated to either of the supersymmetries.
In particular, $J'$ is only associated to a U(3)-structure -- the one in the denominator of the coset description \eqref{cosetSU4} --
since it will induce an integrable complex structure of which the (3,0)-form is globally only well-defined up to a factor.
The existence of these two two-forms, $J'$ and $J$, and the associated integrable and non-integrable complex structures is a generic property of
twistor spaces \cite{standardJ,nonstandardJ}. See also \cite{tomasiellocosets} for a discussion of this in the context of the homogeneous $N=1$ type IIA
solutions. We will construct the two-forms associated to the SU(3)-structures in the next section after we review the conditions these SU(3)-structures have
to satisfy.

\section{Supersymmetry conditions and Bianchi identities}
\label{conditions}

\subsection{Review of the conditions for supersymmetric AdS$_4$ solutions}
\label{condreview}

In this subsection we review the conditions for supersymmetric strict SU(3)-structure AdS$_4$ flux compactifications, first derived in
\cite{lt}. These conditions are equivalent to the set of supersymmetry conditions and Bianchi identities without sources, which as
we mentioned in the introduction suffices to imply all the remaining equations of motion.

It is found that the internal manifold should posses an SU(3)-structure.
This consists of a real two-form $J$ and a complex decomposable three-form $\Omega$
satisfying the compatibility and normalization condition
\subeq{\al{\label{su3compat}
\Omega\wedge J&=0 \, , \\
\label{su3norm}
\Omega\wedge\bar{\Omega} & =\frac{4i}{3}J^3\neq 0 \, ,
}}
and such that the associated metric is positive-definite. Especially the condition for $\Omega$
to be complex decomposable is quite complicated and was studied in \cite{hitchinfunc}. In fact, it was shown
in that paper that this implies that $\Im \Omega$ is (up to a sign) determined by $\Re \Omega$. This works
as follows: from $\Re \Omega$ we can construct first an almost complex structure. We define
\eq{
\label{csnotnor}
\tilde{\mathcal{I}}_k{}^l = \varepsilon^{lm_1\dots m_5} (\Re \Omega)_{km_1m_2} (\Re \Omega)_{m_3m_4m_5} \, ,
}
where $\varepsilon^{m_1\dots m_6}$ is the epsilon-tensor in six dimensions, and
then properly normalize it
\eq{
\label{csnor}
\mathcal{I} = \frac{\tilde{\mathcal{I}}}{\sqrt{-\text{tr}\, \frac{1}{6}\,\tilde{\mathcal{I}}^2}} \, ,
}
so that $\mathcal{I}^2 = -\bbone$.
\eq{
\label{hitchinfunc}
H(\Re \Omega)=\text{tr} \, \frac{1}{6} \, \tilde{\mathcal{I}}^2
}
is called the Hitchin function. This procedure only works if the Hitchin function is strictly negative, which
imposes a condition on $\Re \Omega$. The metric can then be constructed from $\mathcal{I}$ and $J$ through
\eq{
\label{su3metric}
g_{mn}=\mathcal{I}_m{}^lJ_{ln}~,
}
and $\Im \Omega$ is given by
\eq{
\label{ImOm}
\Im \Omega = \mp \frac{1}{3} \mathcal{I}_k{}^l \, \d x^k \wedge \iota_l \Re \Omega \, .
}

Furthermore, the only non-zero SU(3)-structure torsion classes are the scalar $\mathcal{W}_1$
and the primitive (1,1)-form $\mathcal{W}_2$, so that
\subeq{
\label{torsionclasses}
\al{
\label{torsionclass1}
\d J &=-\frac{3}{2}i \, \mathcal{W}_1 \Re \Omega\, , \\
\d \Omega &= \mathcal{W}_1 J\wedge J+\mathcal{W}_2 \wedge J \, .
}}
They can be chosen to be both purely imaginary, and the
sourceless Bianchi identity for $F_2$ imposes moreover
\eq{
\label{c2def}
\d \mathcal{W}_2 = i c_2 \, \Re \Omega \, ,
}
where from the defining properties of $\mathcal{W}_2$ we find for the proportionality factor
\eq{
c_2 = - \frac{1}{8} |{\cal W}^-_2|^2 \, .
}

The warp factor $A$ and the dilaton $\Phi$ are constant, and
the Romans mass is given by
\eq{
\label{romansmass}
\frac{16}{5} e^{2\Phi} m^2 = 3|{\cal W}_1|^2-|{\cal W}_2|^2 \geq 0 \, ,
}
imposing an inequality on the torsion classes.
Finally the RR- and NSNS fluxes are then given by
\subeq{\label{fluxval}\al{
H & =\frac{2m}{5} e^{\Phi}\Re \Omega \, , \label{fluxvalH} \\
F_2 & =\frac{f}{9}J+F'_2 \, , \\
F_4 & =f \mathrm{vol}_4+\frac{3m}{10} J\wedge J \, , \label{21d}
}}
where we have defined
\eq{
\label{parrel}
e^{\Phi} f = \frac{9i}{4} {\cal W}_1  \, , \qquad
e^{\Phi} F'_2 = i {\cal W}_2  \,  .
}

Massaging these conditions a bit and keep only the essential, in the end
we have to solve the following problem. Find a geometry with two-forms $J,\widetilde{\mathcal{W}}_2,L$ and
a three-form $\Re \Omega$ so that
\subeq{\label{allcondsol}\al{
\d J & = c_1 \Re \Omega \, , \label{condsol1} \\
\widetilde{\mathcal{W}}_2 & = \frac{\tilde{c}_2}{c_1} J + L \, , \qquad \d L = 0 \, , \label{W2cond} \\
J \wedge \Re \Omega & = 0 \, , \qquad L \wedge \Re \Omega = 0 \, , \label{condsol3} \\
\Re \Omega\wedge\Im\Omega(\Re\Omega) & = -\frac{2}{3}J^3 \neq 0 \, , \label{normcond} \\
\d \left[ \Im \Omega(\Re\Omega) \right] & = \widetilde{\mathcal{W}}_2 \wedge J \, , \label{condsol5}
}}
where the notation $\Im\Omega(\Re\Omega)$ stresses that $\Im\Omega$
should be found from $\Re\Omega$ through the procedure outlined in
eqs.~(\ref{csnotnor}-\ref{ImOm}), making the conditions containing $\Im \Omega$
the most involved. After we find a solution to the above conditions we must still check for positivity
of the metric. The relation with the parameters introduced earlier is
as follows
\eq{\spl{
c_1 & = - \frac{2}{3} e^{\Phi} f \, , \\
\widetilde{\mathcal{W}}_2 & = \frac{2}{3} c_1 J - i \mathcal{W}_2 \, , \\
\tilde{c}_2 & = \frac{2}{3} (c_1)^2 + c_2 \, , \\
e^{2 \Phi} m^2 & = \frac{5}{4} \left(2 \tilde{c}_2 -(c_1)^2 \right) = \frac{5}{4} \left( 2 \rho -1 \right) (c_1)^2 \ge 0 \, .
}}
We introduced the parameter
\eq{
\label{torsionpar}
\rho = \frac{\tilde{c}_2}{(c_1)^2} \, ,
}
which considering $m^2 \ge 0$ and $c_2 \le 0$ obeys
$\frac{1}{2} \le \rho \le \frac{2}{3}$. The limiting cases correspond to the massless and nearly-K\"ahler case
respectively.

Since we want to construct a solution that respects the SO(4) isometry,
we expand the unknown forms in left-invariant forms, which we build from the
forms of \eqref{leftinvforms} and the extra left-invariant one-form $\d \xi$,
\eq{\label{ansatz}\spl{
J = & k_1(\xi) e^{12} + k_2(\xi) e^{34} + k_3(\xi) \d \xi \wedge e^5 + k_4(\xi) \left[ \cos\theta_k(\xi) (e^{13}+e^{24})+\sin\theta_k(\xi) (e^{14}-e^{23})  \right], \\
\Re \Omega = & u_1(\xi) e^{125} + u_2(\xi) e^{345} + u_3(\xi) \left( e^{145}-e^{235}\right)+u_4(\xi)\left(e^{135}+e^{245} \right)+ \\
& u_5(\xi) \d \xi \wedge e^{12} +
u_6(\xi) \d \xi \wedge e^{34} + u_7(\xi) \d \xi \wedge \left( e^{14}-e^{23} \right) +  u_8(\xi) \d \xi \wedge \left( e^{13}+e^{24} \right) \, \\
L = & l_1(\xi) e^{12} + l_2(\xi) e^{34} + l_3(\xi) \d \xi \wedge e^5 + l_4(\xi) (e^{14}-e^{23}) +l_5(\xi) (e^{13}+e^{24}) \, ,
}}
where the introduction of the angle coordinate $\theta_k(\xi)$ will be convenient later on, and we do not display
$\widetilde{\mathcal{W}}_2$ since it can be trivially expressed in terms of $J$ and $L$ through \eqref{W2cond}.

\begin{sloppypar}Before solving these conditions in general,
we will first use them to find the SU(3)-structures corresponding to the two SO(4)-invariant
supersymmetries of the massless K\"ahler-Einstein geometry.
\end{sloppypar}

\subsection{The massless K\"ahler-Einstein geometry revisited}

In this case we already know the metric explicitly and that such $J$ and $\Re \Omega$, corresponding
to the two supersymmetries invariant under SO(4), exist. We just want to find them in the coordinate
system of section \ref{m0case}. It turns out that to find most of $J$ (i.e.~apart from $\theta_k(\xi)$)
we do not have to solve the full set of conditions \eqref{allcondsol}.
It suffices to plug the ansatz for $J$, \eqref{ansatz}, into \eqref{su3metric}, and further
impose $\mathcal{I}^2 = - \bbone$ (for a proper complex structure) and $J \wedge \d J=0$ (which follows
from \eqref{condsol1} and \eqref{condsol3}). The other conditions then become relatively simple, and in
the end we find for $J$ and $\Omega$:
\eq{\label{SU3KE}\spl{
a^{-1} J = & -\frac{\sin^2 \xi \cos 2\,\xi}{4} e^{12} +  \frac{\cos^2 \xi \cos 2\,\xi}{4} e^{34}
+ \frac{\sin 2\,\xi}{4} \d \xi \wedge e^5 \\
& + \frac{\sin^2 2\,\xi}{8} \left[ \cos \theta_k \left(e^{13}+e^{24}\right)+\sin \theta_k \left(e^{14}-e^{23}\right)\right] \, \\
a^{-3/2} \Omega = & \left( \d \xi - \frac{i}{4} \sin 2\,\xi \, e^5 \right)
\wedge \frac{1}{2}\left[\cos^2 \xi \, (i e^3 +e^4) - \sin^2 \xi \, e^{-i\theta_k}(e^1-i e^2) \right] \\
& \wedge \frac{1}{4}\sin 2 \xi \left[ i e^{i \theta_k} \left(e^1 + i e^2 \right) + e^3 + i e^4\right] \, ,
}}
where $\theta_k(\xi)=\theta_k$ is constant and we chose the upper sign in \eqref{ImOm}. We find furthermore
\eq{
\label{}
c_1 = \frac{4}{\sqrt{a}} \, , \qquad \tilde{c}_2 = \frac{1}{2} (c_1)^2 \, .
}
Note that in general we can get three more solutions by using one or both of
\eq{\spl{
J \rightarrow J \, , \quad \Re \Omega \rightarrow -\Re \Omega \, , \quad \Im \Omega \rightarrow - \Im \Omega, \, \quad c_1 \rightarrow -c_1, \, \quad \tilde{c}_2 \rightarrow \tilde{c}_2 \, , \\
J \rightarrow -J, \, \quad \Re \Omega \rightarrow \Re \Omega, \, \quad \Im \Omega \rightarrow - \Im \Omega, \, \quad c_1 \rightarrow -c_1, \quad \tilde{c}_2 \rightarrow \tilde{c}_2 \label{Jsignchange} \, ,
}}
where the second line corresponds to choosing a different sign in \eqref{ImOm}. In the following we will always choose $c_1 >0$ and the upper
sign in \eqref{ImOm}.

The angle $\theta_k$ corresponds to the freedom of rotating the SU(3)-structure with SO(2)$_R$.
In more detail, applying the transformation \eqref{Rsymmetry} on the solution results in $\theta_k \rightarrow
\theta_k + 2 \phi$. It does not change
the metric, nor, since $m=0$ any of the fluxes\footnote{This is somewhat less obvious for $F_2$, but
can still be checked using (\ref{fluxval},\ref{parrel}).}.
It follows that the solution has two SO(4)-invariant supersymmetries.

Furthermore we remark that the coset $T^{1,-1}$ is invariant under
\eq{
(e^1,e^2) \leftrightarrow (e^3,e^4) \, , \quad e^5 \rightarrow - e^5 \, ,
}
which is completed with
\eq{
\xi \rightarrow \frac{\pi}{2} - \xi \, , \quad \theta_k \rightarrow -\theta_k \, ,
}
into a symmetry of the solution, which, as we will find, extends to the general solution.

Finally, it will be convenient later on to use the freedom of reparameterizing $\xi$ to put
$k_4(\tilde{\xi})=\tilde{\xi}$. For the $m=0$ solution above we find then
\eq{
\label{m0soltilde}
\tilde{\xi}=\frac{a}{8} \sin^2 2\,\xi \, , \quad k_1(\tilde{\xi})k_2(\tilde{\xi}) = \tilde{\xi}^2 - \frac{a}{8} \tilde{\xi} \, , \quad k_1(\tilde{\xi})+k_2(\tilde{\xi}) = \frac{a}{4}-2\tilde{\xi} \, .
}

\subsection{Solving the conditions}
\label{condsol}

In this section, we put ourselves to the task of solving the equations \eqref{allcondsol} for general $1/2 \le \rho \le 2/3$.
As it turns out, the equations \eqref{condsol1}-\eqref{condsol3} are relatively easily to solve and put:
\eq{\spl{
& u_1(\xi) = u_2(\xi)=l_4(\xi)=l_5(\xi)=0 \, , \\
& k_3(\xi) = \frac{\left[k_1(\xi)k_2(\xi) - k_4(\xi)^2\right]'}{k_1(\xi)-k_2(\xi)} \, , \\
& l_3(\xi) = -l'_1(\xi) \, , \quad l_2(\xi)=d_1 - l_1(\xi) \, , \\
& l_1(\xi) = d_1 \frac{k_1(\xi)\left[k_1(\xi)+k_2(\xi)\right]'-\left[k_4(\xi)^2\right]'}{[k_1(\xi)+k_2(\xi)][k_1(\xi)+k_2(\xi)]'- 2[k_4(\xi)^2]'} \, , \\
& u_3(\xi) = -\frac{k_4(\xi) \cos\theta_k(\xi)}{c_1} \, , \quad u_4(\xi) = \frac{k_4(\xi)\sin\theta_k(\xi)}{c_1} \, , \\
& u_5(\xi) = \frac{k_1(\xi)\left[k_1(\xi)+k_2(\xi)\right]'-\left[k_4(\xi)^2\right]'}{c_1[k_1(\xi)-k_2(\xi)]} \, , \quad
u_6(\xi) = -\frac{k_2(\xi)\left[k_1(\xi)+k_2(\xi)\right]'-\left[k_4(\xi)^2\right]'}{c_1[k_1(\xi)-k_2(\xi)]} \, , \\
& u_7(\xi) = \frac{\left[k_4(\xi) \sin\theta_k(\xi)\right]'}{c_1} \, , \quad
u_8(\xi) = \frac{\left[ k_4(\xi) \cos\theta_k(\xi)\right]'}{c_1} \, ,
}}
where prime denotes the derivative to $\xi$.
This leaves four unknown functions $k_1(\xi)$, $k_2(\xi)$, $k_4(\xi)$, $\theta_{k}(\xi)$, and one unknown constant $d_1$.
It turns out that none of the equations in \eqref{allcondsol} puts a constraint on $\theta_k(\xi)$. This follows from
the fact that \eqref{psishift} is a symmetry of the equations and 
the solutions with different $\theta_k(\xi)$ are related to each other by this shift. We will gauge-fix the
non-constant part of this coordinate reparametrization by requiring $g_{\xi\psi}$ to vanish, which forces $\theta_k(\xi)=\theta_k$ to be constant.

To proceed it will be convenient to also fix the reparametrization freedom of $\xi$ by introducing a new coordinate $\tilde{\xi}$ such that
$k_4(\tilde{\xi})=\tilde{\xi}$. Furthermore, if we introduce
\eq{
p(\tilde{\xi}) = \tilde{\xi}^2 - k_1(\tilde{\xi}) k_2(\tilde{\xi}) \, , \quad s(\tilde{\xi})=k_1(\tilde{\xi})+k_2(\tilde{\xi}) \, ,
}
and square the left- and right-hand side of \eqref{normcond} -- removing the square root in \eqref{csnor} --
it takes the following simplified form
\eq{
\label{normcondsimpl}
\frac{\tilde{\xi}^2\left\{[s(\tilde{\xi})-\tilde{\xi}s'(\tilde{\xi})]^2+p(\tilde{\xi})[4-s'(\tilde{\xi})^2]\right\}}{\left\{\frac{1}{2}\left[p(\tilde{\xi})^2\right]'\right\}^2} = (c_1)^4 \, .
}
This is still not easy to solve unless $s'(\tilde{\xi})^2=4$, which is indeed the case for the massless solution \eqref{m0soltilde}.
We will first make this ansatz, and discuss the alternative case afterwards.

\subsubsection*{The ansatz $s'(\tilde{\xi})^2=4$}

More specifically, we put
\eq{
\label{sansatz}
s(\tilde{\xi})=e_1 - 2 \tilde{\xi} \, ,
}
where we made the same choice of sign as in \eqref{m0soltilde}. There is no loss of generality in this sign choice, since it can be changed
using \eqref{Jsignchange}. We can then solve \eqref{normcondsimpl} and find
\eq{
\label{psol}
p(\tilde{\xi})^2 = \frac{e_1}{(c_1)^2} \tilde{\xi}^2 + e_2 \, .
}

Finally we turn to condition \eqref{condsol5}. It should hold when we wedge it with the most general
left-invariant two-form, which contains five degrees of freedom. Because we have already imposed that $\widetilde{\mathcal{W}}_2$ is a (1,1)-form,
it is automatically satisfied when wedging with a (2,0)- or (0,2)-form. This removes two degrees of freedom. One more degree of freedom is eliminated because \eqref{condsol5}
is independent of $\theta_k(\xi)$. So we need only impose
\subeq{\label{resteqs}\al{
& \d \Im \Omega \wedge L = \widetilde{\mathcal{W}}_2 \wedge L \wedge J \Longleftrightarrow \widetilde{\mathcal{W}}_2 \wedge L \wedge J = 0 \, , \\
& \d \Im \Omega \wedge J = \widetilde{\mathcal{W}}_2 \wedge J \wedge J \Longleftrightarrow  \widetilde{\mathcal{W}}_2 \wedge J \wedge J = \frac{2}{3} \, c_1 J^3 \, .
}}
Of course, we found it prudent to check the full \eqref{condsol5} after finding the solutions.

Plugging in \eqref{sansatz} and \eqref{psol} we find in the end two sets of solutions
in terms of $\rho$
\eq{
\label{gensol}
e_2 = 0 \, , \quad  e_1 = \frac{20 \mp 8 \sqrt{4-6\rho}-6 \rho}{(c_1)^2(2+\rho)^2} \, ,
\quad d_1 = \frac{-4 \pm 4 \sqrt{4-6\rho}+6 \rho}{c_1(2+\rho)} \, .
}
Let us now transform back to the original coordinates where $g_{\xi\xi}$ is constant. We need then to solve
for
\eq{
\sqrt{g_{\tilde{\xi}\tilde{\xi}}(\tilde{\xi})} \d \tilde{\xi} = a^{1/2} \d \xi \, .
}
Further normalizing such that $0 \le \xi \le \pi/2$ we find
\eq{
\tilde{\xi}= \frac{c_1 (e_1)^{3/2}}{4\left(c_1 \sqrt{e_1}-1\right)} \sin^2 2 \xi \, , \quad a = \frac{4 \, e_1}{c_1 \sqrt{e_1}-1} \, .
}
The resulting metric is given by
\begin{multline}
\label{genmetric}
a^{-1} \d s^2 =  \d \xi^2 + \frac{\sin^2 \xi}{8} \left[ c_1 \sqrt{e_1} + (2-c_1 \sqrt{e_1}) \cos 2 \xi\right] \left[ (\d \theta_1)^2 + \sin^2 \theta_1 (\d \phi_1)^2 \right] \\
+ \frac{\cos^2 \xi}{8} \left[ c_1 \sqrt{e_1} - (2-c_1 \sqrt{e_1}) \cos 2 \xi\right] \left[ (\d \theta_2)^2 + \sin^2 \theta_2 (\d \phi_2)^2 \right] \\
+ \sin^2 \xi \cos^2 \xi \, \left(\d \psi + \frac{1}{2} \cos \theta_1 \d \phi_1 - \frac{1}{2} \cos \theta_2 \d \phi_2\right)^2 \\
+ \frac{\sin^2 2 \xi}{16}  \left( 2 - c_1 \sqrt{e_1} \right)
\Big[\sin (2 \psi + \theta_k) (\d \theta_1 \d \theta_2 + \sin \theta_1 \sin \theta_2 \d \phi_1 \d \phi_2) \\
 +\cos (2 \psi + \theta_k) (\sin \theta_2 \d \theta_1 \d \phi_2 - \sin \theta_1 \d \theta_2 \d \phi_1)\Big] \, .
\end{multline}

It is easy to check that for $\xi = \epsilon$ and $\xi = \pi/2-\epsilon$ with $\epsilon$ infinitesimal,
the first three lines of the metric take exactly the same form as the metric in the massless K\"ahler-Einstein case in these limits,
while the terms in the last two lines vanish as $\epsilon^2$. The metric is thus regular.

Except for $2-c_1 \sqrt{e_1}=0$, which is the case for the massless K\"ahler-Einstein solution, the metric changes when shifting $\theta_k$.
Moreover the fluxes will also change. So the SO(2)$_R$ symmetry is broken, although it still sends solutions to solutions.

Furthermore, we find that the SU(3)-structure is given by
\eq{\label{genstruc}\spl{
J(\theta_k) = & \frac{1}{4(c_1\sqrt{e_1}-1)} \Big\{2 \, e_1 \sin^2 \xi \left(-2+c_1\sqrt{e_1}-c_1 \sqrt{e_1} \cos 2 \xi \right)  e^{12} \\
& + 2 \, e_1  \cos^2 \xi \left(-2+c_1\sqrt{e_1}+c_1 \sqrt{e_1} \cos 2 \xi \right)  e^{34}
+ 4 \, e_1 \sin 2 \xi \, \d \xi \wedge e^5 \\
& + c_1 (e_1)^{3/2} \sin^2 2 \xi \left[ \cos \theta_k \left( e^{13}+e^{24} \right) + \sin \theta_k \left(e^{14}-e^{23} \right) \right] \Big\} \, , \\
\Omega(\theta_k) = & \frac{(c_1)^3 (e_1)^{3/2}}{8\left(c_1\sqrt{e_1}-1\right)} \, \Omega_{\text{KE}}(\theta_k) \, ,
}}
where $\Omega_{\text{KE}}(\theta_k)$ is the holomorphic three-form of the K\"ahler-Einstein case defined in \eqref{SU3KE}.
The fact that $\Omega$ is proportional to $\Omega_{\text{KE}}$ means that the complex structure is the same on all solutions
with the same $\theta_k$.

\subsubsection*{Beyond the ansatz}

We will now show that there are no further solutions beyond the ansatz $s'(\tilde{\xi})^2=4$.
An important role is played by the regularity of the metric for $\tilde{\xi} \rightarrow 0$. In terms
of the unknown functions $p(\tilde{\xi})$ and $s(\tilde{\xi})$ the relevant components of the metric
are:
\eq{\spl{
& g_{\tilde{\xi}\tilde{\xi}} = \frac{(c_1)^2 p(\tilde{\xi}) p'(\tilde{\xi}){}^2}{\tilde{\xi}^2(-4 \, \tilde{\xi}^2 + s(\tilde{\xi})^2 + 4 \, p(\tilde{\xi}))} \, , \quad
g_{55} = \frac{\tilde{\xi}^2}{(c_1)^2 p(\tilde{\xi})} \, , \\
& g_{11},g_{33} = \frac{\tilde{\xi}}{2 (c_1)^2 p(\tilde{\xi}) p'(\tilde{\xi})} \left[ \left(s(\tilde{\xi}) - \tilde{\xi} s'(\tilde{\xi})\right) \left( s(\tilde{\xi}) \mp \sqrt{-4 \, \tilde{\xi}^2 + s(\tilde{\xi})^2 + 4 \, p(\tilde{\xi})}\right) + 4 \, p(\tilde{\xi}) \right] \, ,
}}
where the upper/lower sign is for $g_{11},g_{33}$ respectively. Requiring that --- after an appropriate coordinate transformation $\xi(\tilde{\xi})$ such that $g_{\xi\xi}=1$ --- the metric takes the form \eqref{metricreg} for $\xi \rightarrow 0$, we find for the
asymptotic behavior of $p(\tilde{\xi})$ and $s(\tilde{\xi})$:
\eq{
p(\tilde{\xi}) = d_1 \tilde{\xi} + \mathcal{O}(\tilde{\xi}^2) \, , \qquad
s(\tilde{\xi}) = d_2 + \mathcal{O}(\tilde{\xi}) \, ,
}
with $(d_2)^2 = (c_1)^4 (d_1)^4$. Solving now eqs.~\eqref{normcondsimpl} and \eqref{resteqs} order by order in $\tilde{\xi}$
we recover solution \eqref{sansatz},\eqref{psol},\eqref{gensol}.

\subsection{Equivalence to the family of $N=1$ massive Sp(2)-invariant solutions}
\label{proofequiv}

In this subsection we show that the family of solutions \eqref{genmetric},\eqref{genstruc} is equivalent
to the family of $N=1$ Sp(2)-invariant solutions of \cite{tomasiellocosets,cosets}. It will be convenient
to use the procedure discussed in section 5.1 of \cite{tomasiellomassive2} to construct the Sp(2)-invariant
solutions in terms of the homogeneous coordinates on $\mathbb{CP}^3$.

As in that paper we start with $\mathbb{C}^4 \,\backslash\, \{ 0\}$, which can be considered as a $\mathbb{C}^*$ bundle over $\mathbb{CP}^3$ with projection map $p$.
We can describe the coordinates $Z^a$ of $\mathbb{C}^4$ in terms of our SO(4)-adapted coordinates as in \eqref{homcoord},
where we take $\lambda = r e^{i\gamma}$. The fiber coordinates are then $(r,\gamma)$.
We are interested in forms $\alpha$ on the total space of the bundle $\mathbb{C}^4 \,\backslash\, \{ 0\}$ that are the pull-back under $p$
of a form on the base space $\mathbb{CP}^3$. Such forms are called {\em basic}. A basic form is both {\em vertical}, i.e.\ $\iota_v \alpha=0$
for all vectors $v$ along the fiber, and {\em invariant}, i.e.\ $\mathcal{L}_v \alpha = 0$. For our bundle, the vectors tangent to the fiber
are spanned by
\eq{
r \partial_r = Z^a \partial_a + \bar{Z}_a \bar{\partial}^a \, , \qquad
\partial_\gamma = i (Z^a \partial_a - \bar{Z}_a \bar{\partial}^{a}) \, ,
}
and the dual one-forms are
\eq{
\frac{\d r}{r} = \frac{1}{2r^2} (\bar{Z}_a \d Z^a + Z^a \d \bar{Z}_a) \, , \qquad \eta = \frac{i}{2r^2} (-\bar{Z}_a \d Z^a + Z^a \d \bar{Z}_a) \, .
}
One can now split a form on $\mathbb{C}^4$ into a vertical and a non-vertical part. For the one-forms
$\d Z^a$ of the complex basis, for instance, one finds:
\eq{
\d Z^a = \left(\delta^a{}_b - \frac{Z^a \bar{Z}_b}{r^2}\right) \d Z^b + Z^a \left(\frac{\d r}{r} +i \eta \right) \, .
}
Using this one finds immediately for the standard K\"ahler form on $\mathbb{C}^4$,
\eq{
J_{(4)} = \frac{i}{2} \d Z^a \wedge \d \bar{Z}_a = r \d r \wedge \eta + r^2 J_{\text{FS}} \, , \qquad J_{\text{FS}} = \frac{i}{2 r^2} D Z^a D \bar{Z}_a \, ,
}
where $J_{\text{FS}}$ is the unit Fubini-Study K\"ahler form. One can indeed check that $J_{\text{FS}}$
is vertical and invariant. Changing to the SO(4)-adapted coordinates \eqref{homcoord} it takes the form $a^{-1} J'$ of \eqref{FSKahl}. Using
$\mathcal{L}_{r \partial_r} J_{(4)}=2 J_{(4)}$, which follows from the fact that $J_{(4)}$ is homogeneous of degree
two in the coordinates $Z^a$, we find furthermore:
\eq{
\label{deta}
\d \eta = J_{\text{FS}} \, .
}
On the other hand, if one decomposes
\eq{
\Omega_{(4)} = \d Z^1 \wedge \d Z^2 \wedge \d Z^3 \wedge \d Z^4 = -i r^3 (\d r + i r \eta) \wedge \Omega_{\text{FS}} \, ,
}
one finds that $\Omega_{\text{FS}}$ is vertical and satisfies $\mathcal{L}_{r \partial_r} \Omega_{\text{FS}}=0$. However, $\Omega_{\text{FS}}$
is not basic since $\mathcal{L}_\gamma \Omega_{\text{FS}} = 4 i \Omega_{\text{FS}} \neq 0$. This is related to the fact that on $\mathbb{CP}^3$
there is no globally defined $(3,0)$-form associated to $J_{\text{FS}}$.

Let us now make a choice of Sp(2) subgroup of SU(4). Such a choice is in one-to-one correspondence with the choice of a {\em holomorphic symplectic
form} $\kappa$. This is a closed two-form, $\d \kappa=0$, whose square gives the holomorphic (4,0)-form
\eq{
\label{kappadef}
\frac{1}{2} \kappa^2 = \Omega_{(4)} \, .
}
Indeed, just like the supersymmetry generators, $\kappa$ transforms as a $\bf{6}$ under SU(4), which decomposes under Sp(2) as $\bf{6} \rightarrow 1 + 5$.
$\kappa$ then corresponds to the invariant part. Expanding $\kappa$ in a vertical and a non-vertical part, one obtains a two-form $t_{\kappa}$
and a one-form $s_{\kappa}$ as follows
\eq{
\label{kappasplit}
\kappa = r (\d r + i r \eta) \wedge s_{\kappa} + r^2 t_{\kappa} \, .
}
By construction, these forms are vertical and one also easily checks their invariance under $r \partial_r$.
They are however {\em not} invariant under $\partial_\gamma$. The simplest way to obtain this, is to first realize
that $\mathcal{L}_{r \partial_r} \kappa = 2 \kappa$, since $\kappa$ is homogeneous of degree two in the complex
coordinates $Z^a$. Together with \eqref{kappasplit} one finds then
\eq{
\d s_{\kappa} = 2 \left( i \eta \wedge s_{\kappa} + t_{\kappa} \right) \, ,
}
and finally, using \eqref{deta},
\eq{
\mathcal{L}_{\gamma} s_{\kappa} = 2 i s_{\kappa} \, , \qquad \mathcal{L}_{\gamma} t_{\kappa} = 2 i t_{\kappa} \, ,
}
so that $s_{\kappa}$ and $t_{\kappa}$ are {\em not} well-defined on $\mathbb{CP}^3$. From \eqref{kappadef} it follows
that, in fact, $\Omega_{\text{FS}}$, which is also not well-defined on $\mathbb{CP}^3$, can be written as
\eq{
\Omega_{\text{FS}} = i s_{\kappa} \wedge t_{\kappa} \, .
}

However, it is now obvious that if we define instead
\eq{
\Omega = i \bar{s}_{\kappa} \wedge t_{\kappa} \, ,
}
the $\gamma$-charges compensate and we do obtain a well-defined (3,0)-form on $\mathbb{CP}^3$. This corresponds
to a different choice of almost complex structure, replacing one one-form of the complex basis by its conjugate. If we want $(J,\Omega)$
to still correspond to the Fubini-Study metric we have to define $J$ by performing the same operation, i.e.\ $s_{\kappa}\leftrightarrow \bar{s}_{\kappa}$, on $J_{\text{FS}}$. This amounts to\footnote{The overall sign in
the definition of $J$ ensures that $J^3=J_{\text{FS}}^3$.}
\eq{
J = -\left(J_{\text{FS}} - i s_{\kappa} \wedge \bar{s}_{\kappa}\right) \, .
}
In fact, generalizing this construction we can define a whole family of different SU(3)-structures $(J_{\sigma},\Omega_{\sigma})$,
corresponding to different metrics on $\mathbb{CP}^3$, as follows:
\eq{\spl{
J_{\sigma} & = - a \left( \frac{2}{\sigma} J_{\text{FS}} - i \, \frac{\sigma+2}{2 \sigma} \, s_{\kappa} \wedge \bar{s}_{\kappa} \right) \, , \\
 \Omega_{\sigma} & = \frac{2ia^{3/2}}{\sigma} \, \bar{s}_{\kappa} \wedge t_{\kappa} \, ,
}}
where $a>0$ is an overall scale and $\sigma$ is a shape parameter.
One can check that these $(J_{\sigma},\Omega_{\sigma})$ solve the supersymmetry conditions \eqref{torsionclasses} and \eqref{c2def}
with
\eq{
-i \mathcal{W}_1 = \frac{2\, c_1}{3} =\frac{2(2+\sigma)}{3 a^{1/2}} \, , \qquad |\mathcal{W}_2|^2 = -8 c_2 = \frac{64(1-\sigma)^2}{3 a} \, .
}
The condition \eqref{romansmass} is satisfied for $2/5 \le \sigma \le 2$, and the fluxes can be found from \eqref{fluxval} and \eqref{parrel}.
The parameter $\rho$ of eq.~\eqref{torsionpar} is given by
\eq{
\label{rhoval}
\rho = \frac{2 \sigma (4-\sigma)}{(2+\sigma)^2} \, .
}
Since these solutions also manifestly preserve the Sp(2) singled out by $\kappa$, they correspond in fact to the $N=1$
family of solutions first introduced in \cite{tomasiellocosets}.

From the discussion of section \ref{symmetries} it follows that to make contact with our solutions we should take
\eq{
\label{choicekappa}
\kappa = e^{i \theta/2}\d Z^1 \wedge \d Z^2 + e^{-i \theta/2}\d Z^3 \wedge \d Z^4 \, .
}
Indeed, our specific choice of SO(4), of which the action was defined around \eqref{so4action}, leaves
both terms in \eqref{choicekappa} separately invariant, corresponding to $\bf{6}\rightarrow (\bf{1},\bf{1}) \oplus (\bf{1},\bf{1}) \oplus (\bf{2},\bf{1}) \oplus (\bf{1},\bf{2})$. So any choice of $\theta$ corresponds
to a choice of Sp(2) so that SO(4)$\subset$Sp(2). Going through the above construction,
we find that $(J_{\sigma},\Omega_{\sigma})$ exactly matches \eqref{genstruc} upon putting
\eq{
\label{e1val}
e_1 = \frac{a}{4} \, , \qquad \theta_k = \theta - \frac{\pi}{2} \, .
}
The transformation \eqref{Rsymmetry} corresponds to shifting $\theta \rightarrow \theta + 2 \phi$.
It will rotate the SU(3)-structure $(J_{\sigma},\Omega_{\sigma})$ and thus supersymmetry generator
as well as the Sp(2) symmetry group.
In the massless K\"ahler-Einstein case, $\sigma=2$, it leaves the metric and the fluxes invariant so that it really
corresponds to an R-symmetry, while in all other cases it rotates a solution into a different solution.

We conclude that the solutions of \eqref{genstruc} enhance their symmetry group to Sp(2)
and form the $N=1$ family of homogeneous solutions first introduced in \cite{tomasiellocosets}.
Moreover, we have shown that there are no other solutions with SO(4) symmetry group and strict SU(3)-structure.

\section{Coisotropic D-brane embeddings}
\label{coisemb}

Now that we have constructed the SU(3)-structures associated to the supersymmetry in an SO(4)-invariant
description, let us see whether the geometry allows for the embedding of supersymmetric D-branes,
which according to \cite{gencal,lucacal} must be generalized calibrated.
It is known \cite{slagCP3,slagCP32,slagCP33} that there are special Lagrangian D-branes wrapping $\mathbb{RP}^3$s, but we
will here be interested in the more exotic coisotropic D8-branes \cite{coisotropic}.

A coisotropic space-time-filling D8-brane has one transversal direction in $\mathbb{CP}^3$ and is thus a priori compatible with the
symmetry group SO(4), acting with cohomogeneity one. So let us construct an SO(4)-invariant coisotropic D8-brane.
To be left-invariant, the one-form part of the D-brane source $j$, defining the direction
transversal to the D-brane, must be a linear combination of the one-forms $\d \xi$ and $e^5$.
In fact, the only coordinate dependence allowed by the SO(4) symmetry group is the dependence on $\xi$, so
if we want a localized D-brane, we must take the transversal coordinate to be $\xi$.
The coisotropic D-brane then wraps $T^{1,-1}$ at some constant value $\xi=\xi_0$.
For the world-volume gauge field $\mathcal{F}$ we make the SO(4)-invariant ansatz
\eq{
\label{Fansatz}
\mathcal{F} = f_1 e^{12} + f_2 e^{34} + f_3 \left( e^{14}-e^{23} \right) +f_4 \left( e^{13} + e^{24} \right) \, ,
}
with $f_i$ constant. It must satisfy
\eq{
\d \mathcal{F} = H|_\Sigma \, .
}

The generalized calibration condition \cite{gencal,lucacal} for a space-time-filling D8-brane
becomes
\eq{
\label{calcond}
\left(i J|_\Sigma + \mathcal{F}\right)^2=0 \, , \qquad \Re \Omega|_\Sigma \wedge \mathcal{F}=0 \, .
}
The first condition, which can be interpreted as an F-flatness condition in the low-energy effective theory \cite{lucasuppot},
is equivalent to the condition for a coisotropic D-brane \cite{coisotropic}, see e.g.~\cite{coisotropicmarchesano,deforms}. This has to be completed
with the second condition, which can be interpreted as a D-flatness condition. However, in \cite{DbraneAdS} it was shown that in AdS$_4$
compactifications the second condition follows automatically from the first.

Let us first consider the massless K\"ahler-Einstein background. In this case, we find from $H=0$ that $\mathcal{F}$
must be closed, implying $f_3=f_4=0$.
Furthermore, we find from \eqref{calcond}
\eq{
\label{coism0}
\xi_0 = \pi/4, \quad 64 \, f_1 f_2=-a^2 \, , \quad \text{or} \quad \xi_0=0, \, \quad f_1=0 \, , \quad \text{or} \quad \xi_0=\pi/2, \, \quad f_2=0 \, ,
}
Since \eqref{calcond} then holds no matter the choice of $\theta$ in \eqref{SU3KE}, the D-brane preserves both SO(4)-invariant supersymmetries.
Only the first possibility leads to a genuine D8-brane, since for the last two possibilities the second respectively the
first $\mathbb{CP}^1$ and the fiber shrink to zero and the D-brane only wraps the other $\mathbb{CP}^1$.

In the full quantum theory, the closed world-volume gauge field $F_{\text{WV}} = \mathcal{F} - B|_{\Sigma}$ must be quantized
\eq{
\label{fluxquant}
\frac{1}{(l_s)^2} \int_{S^2} F_{\text{WV}} = n_F \, ,
}
where $l_s= 2 \pi \sqrt{\alpha'}$ is the string length, $S^2$ is a representative of the two-cycle homology of $T^{1,-1}$
and $n_F$ is an integer. We can take $B|_{\Sigma}=0$. This will boil down to
\eq{
\frac{a}{(l_s)^2} d (\tilde{f}_1 + \tilde{f}_2) = n_F \, ,
}
where $d$ is a coefficient of order one (depending on the details of the integral) and we introduced the dimensionless $\tilde{f}_i = a^{-1} f_i$.
There is no problem in choosing the $\tilde{f}_i$ such that $n_F$ is integer and $64 \, \tilde{f}_1 \tilde{f}_2 = -1$ (which follows from \eqref{coism0}).

We now move on to the general massive $N=1$ background \eqref{genstruc}, where we put \eqref{rhoval} and \eqref{e1val}.
This time, since $H \neq 0$ the condition on the exterior derivative
of $\mathcal{F}$ is more complicated:
\eq{
\d \mathcal{F} = H|_\Sigma = \frac{2m}{5} e^{\Phi} \Re \Omega|_{\Sigma} = \sqrt{\frac{-5 \sigma^2 + 12 \sigma -4}{5a}} \, \Re \Omega|_{\Sigma} \, ,
}
where we used \eqref{fluxvalH} and \eqref{romansmass}. The
calibration conditions are still the same as in \eqref{calcond}. We find the following solution:
\eq{\spl{
f_1 & = a \sin^2 \xi_0 \frac{\sigma (2 + \sigma) \sqrt{-5 \sigma^2+12 \sigma-4} \, \sin^2 2\,\xi_0 \mp 8  \sqrt{ \sigma^3 \left[ 1 - 3\sigma -(1+2\sigma) \cos 4\,\xi_0 \right]}}{8 \sqrt{5}\sigma^2  \left[-2 + \sigma + (2+\sigma) \cos 2 \, \xi_0\right]} \, , \\
f_2 & = a \cos^2 \xi_0 \frac{\sigma (2 + \sigma) \sqrt{-5 \sigma^2+12 \sigma-4} \, \sin^2 2\,\xi_0 \pm 8 \sqrt{ \sigma^3 \left[ 1 - 3\sigma -(1+2\sigma) \cos 4\,\xi_0 \right]}}{8 \sqrt{5}\sigma^2  \left[-2 + \sigma - (2+\sigma) \cos 2\,\xi_0\right]} \, , \\
f_3 & = a\frac{\sqrt{-5 \sigma^2+12 \sigma-4}}{16 \sqrt{5}\sigma} \sin^2 2\,\xi_0 \sin \theta_k \, , \\
f_4 & = a\frac{\sqrt{-5 \sigma^2+12 \sigma-4}}{16 \sqrt{5}\sigma} \sin^2 2\,\xi_0 \cos \theta_k \, , \\
}}
where the D-brane can be at any $\xi=\xi_0$ such that $1 - 3\sigma -(1+2\sigma) \cos 4 \xi_0 \ge 0$ or
$\frac{1}{4} \arccos \left(\frac{1-3 \sigma}{1+ 2\sigma} \right) \le \xi_0 \le \frac{\pi}{2} - \frac{1}{4} \arccos \left( \frac{1-3 \sigma}{1+ 2\sigma} \right)$. The D-brane then preserves the $N=1$ supersymmetry of the background. In the K\"ahler-Einstein limit, $\sigma \rightarrow 2$, the allowed interval shrinks to just the point $\xi_0=\frac{\pi}{4}$ and we obtain the $N=2$ solution
above. Moving towards the squashed massless solution, $\sigma \rightarrow 2/5$, on the other hand, the interval gets larger and becomes
$\frac{1}{4} \arccos \left( -\frac{1}{9} \right) \le \xi_0 \le \frac{\pi}{2} - \frac{1}{4} \arccos \left( -\frac{1}{9} \right)$ at this point.

Using the freedom in $\xi_0$, again solutions can be found such that the world-volume gauge field is quantized as in \eqref{fluxquant}.
This time the $B$-field is non-zero, but it can still be globally defined. We can take $B = \frac{2m}{5 c_1} \Re \Omega$. This
$B$-field also plays an important role in the flux quantization of the bulk RR-charges (for a discussion see \cite{tomasiellocosets,cassred}). Indeed, the RR-fields have to be twisted with $e^B$ in order to obtain closed forms with well-defined periods.

\section{Concluding remarks}
\label{outlook}

In this paper I have shown that the only solutions on AdS$_4\times\mathbb{CP}^3$
with symmetry group SO(4) and a strict SU(3)-structure ansatz for the supersymmetry, are
the homogeneous $N=1$ solutions of \cite{tomasiellocosets}, which possess the larger symmetry
group Sp(2). This is in agreement with the results of \cite{tomasiellomassive}, where although a Chern-Simons-matter
theory with SO(4) global symmetry was proposed, it was also argued that the geometric dual should
have SU(3)$\times$SU(3)-structure instead. In the meantime, in \cite{tomasiellomassive2} this dual has been constructed to first
order in the Romans mass parameter $m$, making use of the explicit form of the D2-brane superpotential, obtained
form the CFT. This is particularly interesting since, so far, there were no examples of source-less supergravity compactifications with dynamic SU(3)$\times$SU(3)-structure (see \cite{petrinimassive} for another recent construction).
The construction of the full solution seems more difficult, since the SU(3)$\times$SU(3) supersymmetry
conditions are more complicated than the ones for the strict SU(3)-structure ansatz. The basic procedure should however be the
same as in this paper, namely to expand the pure spinors defining the SU(3)$\times$SU(3)-structure in SO(4)
left-invariant forms and try to solve the differential equations in one variable.
As in \cite{tomasiellomassive2} one could also insert the extra data of the D2-brane superpotential.

Furthermore, I have presented probe supersymmetric D8-branes of the coisotropic type on both the K\"ahler-Einstein ABJM geometry
as well as the massive Sp(2) family.  It would be interesting to find the dual CFT that would correspond to adding
D8-branes in such a way. Presumably, as is the case for adding special Lagrangian D6-branes \cite{slagCP3,slagCP32,slagCP33}
this would add flavors to the Chern-Simons-matter theory.

Finally, one could also try to construct similar solutions on the $\frac{\text{SU(3)}}{\text{U(1)}\times \text{U(1)}}$ manifold, for which e.g.\ \cite{tomasielloN3} could provide a starting point.

\begin{acknowledgments}
I would like to thank Claudio Caviezel, Johanna Erdmenger, Simon K\"ors and Luca Martucci for useful discussions.
Furthermore I thank Dario Martelli and James Sparks for pointing out a fatal flaw in the first version, and
Davide Gaiotto and Alessandro Tomasiello for suggesting an approach to rewrite the paper.
For many of the calculations I found it convenient to use Mathematica together with the package {\tt scalarEDCcode}
for handling forms ({\tt http://library.wolfram.com/infocenter/MathSource/683}).

I am supported by the German Research Foundation (DFG) within the Emmy-Noether-Program (Grant number ZA 279/1-2).
\end{acknowledgments}

\appendix

\bibliographystyle{JHEP}
\bibliography{cosetnotes}
\end{document}